\documentclass[twocolumn,amsmath,amssymb,pra,a4paper]{revtex4-1}
\usepackage{graphicx}
\usepackage[paperwidth=212mm,paperheight=297mm,centering,hmargin=1.65cm,vmargin=2.6cm]{geometry}
\usepackage{graphicx}
\usepackage{dcolumn}
\usepackage{bm}
\usepackage{upgreek}
\usepackage{amsmath}
\usepackage{url}
\usepackage{hyperref}
\usepackage{lipsum}

\newcommand{\beq}{\begin{equation}}
\newcommand{\eeq}{\end{equation}}
\newcommand{\bea}{\begin{align}}
\newcommand{\eea}{\end{align}}
\newcommand{\beqa}{\begin{eqnarray}}
\newcommand{\eeqa}{\end{eqnarray}}
\newcommand{\e}{\mathrm{e}}
\newcommand{\w}{\omega}
\newcommand{\ket}[1]{\left| #1 \right\rangle}

\newcommand{\av}[1]{\langle #1\rangle}

\newcommand{\ketbra}[2]{\left|#1\right\rangle\hskip-1mm\left\langle #2\right|}

\newcommand{\TU}{\tilde{\Upsilon}(t,\tau)}
\newcommand{\Ltau}{\mathcal{L}(\tau)}

\begin{document}
\title{Optical Signatures of Non-Markovian Behaviour in Open Quantum Systems}
\author{Dara P. S. McCutcheon}
\affiliation{Department of Photonics Engineering, DTU Fotonik, {\O}rsteds Plads, 2800 Kongens Lyngby, Denmark}

\date{\today}

\begin{abstract}

We derive an extension to the quantum regression theorem which facilitates 
the calculation of two-time correlation functions and emission spectra for systems 
undergoing non-Markovian evolution. 
The derivation exploits projection operator techniques, with which we obtain  
explicit equations of motion for the correlation functions, making only a second order expansion 
in the system--environment coupling strength, and invoking the Born approximation at a fixed initial time. 
The results are used to investigate a driven semiconductor quantum dot coupled to an acoustic phonon bath, 
where we find the non-Markovian nature of the dynamics has observable signatures in the form
of phonon sidebands in the resonance fluorescence emission spectrum. Furthermore, we use recently 
developed non-Markovianity measures to demonstrate an associated flow of information from the phonon bath 
back into the quantum dot exciton system.

\end{abstract}
\maketitle

\section{Introduction}

Two-time correlation functions are quantities of frequent interest in many areas of physics. 
This is particularly true in quantum optics, where correlation functions  
of the form $\av{A(t_1) B(t_2)}$ give the field correlation properties of an emitting system such as a driven atom, and whose 
Fourier transform gives the measured spectrum~\cite{mollow69}. 
If the governing Hamiltonian can be diagonalised exactly, 
calculation of the two-time correlation function is no more challenging than calculating a one-time expectation value of the form 
$\av{A(t_1)}$. However, it is more often the case that the emitting system is an open system, whose dynamics 
can only be approximated. In this case, since the system operators 
$A$ and $B$ are evaluated at two distinct times, calculation of the correlation function given knowledge of 
system dynamics alone is not at first sight straightforward. The quantum regression theorem, however, gives a prescription   
of how such correlation functions can be related to more readily obtainable system expectation values~\cite{carmichael}. 
A subtle caveat of the quantum regression theorem, however, is that it applies only to systems undergoing strictly Markovian evolution. 
It requires that the complete density operator of the system and environment factorises at all times, and that 
the reduced system density operator obeys a time-independent master 
equation~\cite{Swain1999,Alonso95,DeVega2008,Flindt2008,Budini2008,Goan2010,Goan2011,Guarnieri2014}.

The requirement of Markovian evolution is typically fulfilled in the traditional case of atomic quantum optics 
due to the extremely short correlation time of the electromagnetic environment~\cite{mandelbook,Koshino2004}. 
However, more recent technological advances in the fabrication of artificial emitters 
and the engineering of structured environments have given rise to systems whose evolution is not purely Markovian, yet whose 
properties are typically probed optically. These systems include semiconductor quantum dots (QDs), 
for which Rabi oscillations~\cite{flagg09,ramsay10,monniello13}, resonance fluorescence~\cite{ates09,ulrich11_short,Wei14,matthiesen12,Matthiesen2013}, 
and single photon emission~\cite{michler00,santori02,flagg10} have all been demonstrated. QDs, however, exist in a solid-state 
substrate, and interactions with phonons and nuclear spins can modify their emission 
properties~\cite{mccutcheon13,monniello13,roy11,Roy-choudhury2015} and also 
give rise to non-Markovian behaviour~\cite{mccutcheon10_2,Kaer2014,luka09,barnes12,Ubbelohde2012}. Additionally, for 
technological applications, such as indistinguishable and entangled photon sources~\cite{Lindner2009,gazzano13,Muller2014,McCutcheon2014}, 
it is often desirable to place artificial emitters in structured photonic environments such as in photonic crystals or micro-pillar cavities, 
which also have the potential to lead to non-Markovian behaviour. 

Thus, in order to model the optical properties of these ever more exotic systems, it is important 
to establish how two-time correlation functions can be calculated for open systems undergoing non-Markovian evolution.
We note that efforts in this direction have been 
made~\cite{Swain1999,Alonso95,DeVega2008,Flindt2008,Budini2008,Goan2010,Goan2011,Guarnieri2014}, 
and the conditions under which 
the regression theorem holds have been scrutinised~\cite{Budini2008}. 
Many of these, however, 
rely on a number of uncontrolled 
approximations, such as artificially enforcing time-locality~\cite{Goan2010,Goan2011}, 
or assuming a restrictive (rotating wave-like) form of the system--environment coupling~\cite{Alonso95,DeVega2008}. 
Additionally, it is not clear to what extent non-Markovian behaviour 
has any measurable optical consequences in physically relevant systems. 

In this work we use projection operator techniques to derive a 
non-Markovian extension to the quantum regression theorem, 
valid to second order in the system--environment coupling strength, 
and invoking the Born approximation only at a single fixed initial time. 
The second order expansion restricts the theory to weak--system environment coupling regimes 
for which non-Markovian behaviour is typically only present for short times, and which is usually 
very challenging to observe. The key advantage of the present work, however, 
is that this short-time behaviour is of a two-time correlation function, whose spectral counterpart 
corresponds to a concrete readily measurable quantity. 
Specifically, we apply our formalism to the relevant case of a driven 
QD~\cite{ates09,ulrich11_short,Wei14,matthiesen12,Matthiesen2013}, and find that the 
experimentally observed phonon sidebands in the emission spectra are a direct consequence of non-Markovian behaviour, 
which the standard Markovian treatment fails to capture. 
Moreover, we confirm true non-Markovianity and indivisibility of the underlying dynamical 
map by demonstrating that the phonon sidebands are associated with a flow of information from the 
phonon environment back into the QD system~\cite{breuer09}.

\section{Two-time correlation functions and the Regression Theorem}

We begin by introducing two-time correlation functions and the standard (Markovian)
regression theorem. 
We consider a system $S$ interacting with an environment $E$, and wish 
to calculate two-time correlation functions of the form 
$G(t,\tau)=\av{A(t+\tau) B(t)}=\mathrm{Tr}_{S+E}\big[A(t+\tau) B(t) \chi(0)\big]$, 
where $A$ and $B$ are system operators, 
$\chi(0)$ is the total system-plus-environment state at $t=0$, and $\mathrm{Tr}_{S+E}$ denotes a 
trace over both $S$ and $E$. 
For a time independent Hamiltonian $H$ we have 
$A(t)=U^{\dagger}(t) A U(t)$ with $U(t)=\exp[-i H t]$ (we set $\hbar=1$), 
and using the cyclic property of the trace we find 
\begin{align}
G(t,\tau)&=\mathrm{Tr}_S\big[A \,\Lambda(t,\tau)\big],\label{GALambda}\\
\intertext{where the system operator $\Lambda(t,\tau)$ is given by}
\Lambda(t,\tau)&=\mathrm{Tr}_E\big[U(\tau)B \chi(t) U^{\dagger}(\tau)\big],
\label{LambdaDefinition}
\end{align}
with $\chi(t)=U(t)\chi(0)U^{\dagger}(t)$. From Eq.~({\ref{GALambda}}) 
we see that calculation of $G(t,\tau)$ amounts to calculating something analogous to the expectation 
value of $A$, but with respect to the operator $\Lambda(t,\tau)$ rather than the reduced system density 
operator $\rho(t)=\mathrm{Tr}_E[U(t)\chi(0)U^{\dagger}(t)]$.
For this reason we refer to $\Lambda(t,\tau)$ as the reduced \emph{effective density operator}, and 
$\rho(t)$ the reduced \emph{physical density operator}.

The standard regression theorem proceeds by observing that the definition of the effective density operator $\Lambda(t,\tau)$ in 
Eq.~({\ref{LambdaDefinition}}) bares a strong 
resemblance to that of the reduced physical density operator, $\rho(t)=\mathrm{Tr}_E[U(t)\chi(0)U^{\dagger}(t)]$. As such, if we 
know the equation of motion for the physical density operator with respect to $t$, say $\partial_{t}\rho(t)=\Phi\rho(t)$, 
then the reduced effective density operator will obey the same equation of motion but with respect to $\tau$, and with a modified 
initial condition, namely $\partial_{\tau}\Lambda(t,\tau)=\Phi\Lambda(t,\tau)$ and $\Lambda(t,0)=B\rho(t)$. 
We will see, however, that this procedure contains a hidden assumption that the total physical density operator $\chi(t)$ factorises 
for all times~\cite{Swain1999,Alonso95,DeVega2008,Goan2010}.

\subsection{Effective Density Operator Master Equation Using Projection Operators}

To see how this assumption arises, and how it can be removed, we 
now derive the quantum regression theorem 
using the projection operator formalism~\cite{b+p,nakajima58,zwanzig60,shibata77}. 
This well-established formalism was originally developed to calculate physical density operator master equations, 
and our purpose here is to do the same for the effective density operator, taking particular care to identify places where 
any approximations have different physical significance. 
To begin we must establish an interaction picture for the total effective density operator, which we define 
as $\Upsilon(t,\tau)=U(\tau)B\chi(t)U^{\dagger}(\tau)$, such that $\smash{\Lambda(t,\tau)=\mathrm{Tr}_E[\Upsilon(t,\tau)]}$. We write 
the total Hamiltonian $H=H_0+\alpha H_I$, where $H_0=H_S+H_E$ with $H_S$ and $H_E$ acting exclusively on  
$S$ and $E$ respectively. 
We recall that the unitary operators $U(\tau)$ and $U_0(\tau)$ are defined as the solutions to the differential equations 
$i\partial_{\tau} U(\tau)=H U(\tau)$ and $i\partial_{\tau} U_0(\tau)=H_0 U_0(\tau)$, and the interaction picture effective density operator 
as $\tilde{\Upsilon}(t,\tau)=U_I(\tau) B \chi(t) U_I^{\dagger}(\tau)$ with 
$U_I(\tau)=U_0^{\dagger}(\tau)U(\tau)$. From these definitions we find 
\beq
\partial_\tau \tilde{\Upsilon}(t,\tau)=-i\alpha [\tilde{H}_I(\tau),\tilde{\Upsilon}(t,\tau)]=\alpha\mathcal{L}(\tau)\tilde{\Upsilon}(t,\tau),
\label{LDefinition}
\eeq
where $\tilde{H}_I(\tau)=U_0^{\dagger}(\tau) H_I U_0(\tau)$ and the 
Liouvillian $\mathcal{L}(\tau)$ is defined to satisfy the second equality. We naturally 
define $\tilde{\Lambda}(t,\tau)=\mathrm{Tr}_E[\tilde{\Upsilon}(t,\tau)]$, and note that since we can write 
$U_0(\tau)=U_S(\tau)U_E(\tau)$ 
with the subscripts indicating whether the operators act on $S$ or $E$ 
we find $\Lambda(t,\tau)=U_S(\tau)\tilde{\Lambda}(t,\tau) U_S^{\dagger}(\tau)$. The Schr\"{o}dinger 
and interaction picture equations of motion are then related through 
\beq
\partial_{\tau}\Lambda(t,\tau)=i[\Lambda(t,\tau),H_S]+U_S(\tau)\big(\partial_{\tau}\tilde{\Lambda}(t,\tau)\big)U_S^{\dagger}(\tau).
\eeq
These results demonstrate that the effective density operator has a well-defined interaction picture 
which facilitates the use of the master equation techniques below.

We now introduce the projection operators $\mathcal{P}$ and $\mathcal{Q}=(\openone-\mathcal{P})$, 
which are defined through~\cite{nakajima58,zwanzig60,shibata77}
\begin{align}
\mathcal{P}\tilde{\Upsilon}(t,\tau)=\mathrm{Tr}_E[\tilde{\Upsilon}(t,\tau)]\otimes\rho_R=\tilde{\Lambda}(t,\tau)\otimes\rho_R,
\end{align}
where $\rho_R$ is a reference state of the environment. The projection operators 
project the effective density operator into 
factorising and non-factorising components, i.e. we can write $\tilde{\Upsilon}(t,\tau)=(\mathcal{P}+\mathcal{Q})\tilde{\Upsilon}(t,\tau)$, 
where the first term factorises by definition, while the second captures those components which do not. From these basic definitions 
one can show that $\mathcal{P}^2=\mathcal{P}$ and $\mathcal{Q}^2=\mathcal{Q}$, while 
$\mathcal{Q}\mathcal{P}=\mathcal{P}\mathcal{Q}=0$. 
In what follows we assume $\mathrm{Tr}_E[ H_I\rho_R]=0$.  
This is not an approximation, since if $\mathrm{Tr}_E[H_I \rho_R]=\av{H_I}\neq 0$
we can redefine $H_S'=H_S+\av{H_I}$ and $H_I'=H_I-\av{H_I}$ leaving the total Hamiltonian unchanged, and 
we then have $\mathrm{Tr}_E [H_I' \rho_E]=0$ by definition~\cite{mccutcheon11_2,Jang2009}. 
Provided our reference state is chosen 
such that $[H_E,\rho_E]=0$, valid for e.g. thermal states, 
we find $\mathrm{Tr}_E[ \tilde{H}_I'(\tau)\rho_R]=0$ which implies $\mathcal{P}\mathcal{L}(\tau)\mathcal{P}=0$. 

Now, our aim is to derive an equation of motion for the factorising part of the effective density operator $\mathcal{P}\tilde{\Upsilon}(t,\tau)$, 
from which we can readily obtain $\Lambda(t,\tau)=\mathrm{Tr}_E[\mathcal{P}\tilde{\Upsilon}(t,\tau)]$, 
and using Eq.~({\ref{GALambda}}) calculate the two-time correlation function. 
Following Ref.~\cite{b+p} we 
act with both $\mathcal{P}$ and $\mathcal{Q}$ on Eq.~({\ref{LDefinition}}) yielding two differential 
equations which we must solve simultaneously. Inserting $\openone=\mathcal{P}+\mathcal{Q}$ on the right hand side and 
using $\mathcal{P}\mathcal{L}(\tau)\mathcal{P}=0$ the first of these becomes
\beq
\partial_{\tau}\mathcal{P}\TU = \alpha \mathcal{P}\mathcal{L}(\tau)\mathcal{Q}\TU,
\label{PU1}
\eeq
while the second involving $\partial_{\tau}\mathcal{Q}\tilde{\Upsilon}(t,\tau)$ 
can be formally integrated to give
\begin{align}
\mathcal{Q}\TU=&\,\,G_F(\tau,0)\mathcal{Q}\tilde{\Upsilon}(t,0)\nonumber\\
+&\,\,\alpha \int_0^{\tau}\mathrm{d}s G_F(\tau,s)\mathcal{Q}\mathcal{L}(s)\mathcal{P}\tilde{\Upsilon}(t,s),
\label{solQ}
\end{align}
where $G_F(\tau,s)=\mathrm{T}_{\leftarrow}\exp\left[\alpha\int_s^{\tau}\mathrm{d}s'\mathcal{Q}\mathcal{L}(s')\right]$ 
with $\mathrm{T}_{\leftarrow}$ the chronological time ordering operator~\cite{b+p}. 
To obtain a time-local form, from Eq.~({\ref{LDefinition}}) we see that we can write 
$\tilde{\Upsilon}(t,s)=G_B(\tau,s)\TU$, where 
$G_B(\tau,s)=\mathrm{T}_{\rightarrow}\exp\left[-\alpha\int_s^{\tau}\mathrm{d}s'\mathcal{L}(s')\right]$ 
with $\mathrm{T}_{\rightarrow}$ the anti-chronological time ordering operator. 
From Eq.~({\ref{solQ}}) we then find 
\beq
(\openone-\Sigma(\tau))\mathcal{Q}\TU = G_F(\tau,0)\mathcal{Q}\tilde{\Upsilon}(t,0) + \Sigma(\tau)\mathcal{P}\TU,
\label{QTU}
\eeq
where $\Sigma(\tau)=\alpha\int_0^{\tau}\mathrm{d}sG_F(\tau,s)\mathcal{Q}\mathcal{L}(s)\mathcal{P}G_B(\tau,s)$. 
Provided the inverse of the operator $(\openone-\Sigma(\tau))$ exists, Eq.~({\ref{QTU}}) can be solved for 
$\mathcal{Q}\TU$. Since we are ultimately 
interested in the weak-coupling limit of the system--environment interaction $\alpha$, and since 
$\Sigma(\tau)$ contains no zeroth order term in $\alpha$, we assume the existence of 
such an operator, and in solving for $\mathcal{Q}\TU$ we obtain 
\begin{align}
\mathcal{Q}\TU=&\,(\openone-\Sigma(\tau))^{-1}\Sigma(\tau)\mathcal{P}\TU\nonumber\\
+&\,(\openone-\Sigma(\tau))^{-1}G_F(\tau,0)\mathcal{Q}\tilde{\Upsilon}(t,0).
\label{solQUpsilon}
\end{align}
Inserting this formal solution for the non-factorising component of the effective density operator 
into Eq.~({\ref{PU1}}) for the factorising component we find 
\beq
\partial_{\tau}\mathcal{P}\TU=\mathcal{I}(\tau)\mathcal{Q}\tilde{\Upsilon}(t,0)+\mathcal{K}(\tau)\mathcal{P}\TU,
\label{derPExact}
\eeq
where we have defined the kernels 
\begin{align}
\mathcal{I}(\tau)&=\alpha \mathcal{P}\Ltau(\openone-\Sigma(\tau))^{-1}G_F(\tau,0)\mathcal{Q},\label{IDefinition}\\
\mathcal{K}(\tau)&=\alpha \mathcal{P}\Ltau(\openone-\Sigma(\tau))^{-1}\Sigma(\tau)\mathcal{P}\label{KDefinition}.
\end{align}
These expressions constitute an exact equation of motion for the reduced effective density operator, 
with an inhomogeneous term which depends on the physical density operator through 
$\mathcal{Q}\tilde{\Upsilon}(t,0)=\mathcal{Q} B \chi(t)$.

For these reasons, it what follows it will be useful to also consider the evolution  
for the factorising and non-factorising parts of the physical density operator $\chi(t)$. 
For this purpose we use the projection operator methods outlined above, 
and we find that the derivation proceeds in precisely the same manner, 
the only difference being that the time argument $\tau$ is 
replaced with $t$ and the initial condition is $(\mathcal{P}+\mathcal{Q})\chi(0)$. 
In exact analogy with Eq.~({\ref{solQUpsilon}}), we find that the non-factorising part has solution
\begin{align}
\mathcal{Q}\tilde{\chi}(t)=&\,(\openone-\Sigma(t))^{-1}\Sigma(t)\mathcal{P}\tilde{\chi}(t)\nonumber\\
+&\,(\openone-\Sigma(t))^{-1}G_F(t,0)\mathcal{Q}\tilde{\chi}(0),
\label{solQchi}
\end{align}
leading to the equation of motion 
\beq
\partial_{t}\mathcal{P}\tilde{\chi}(t)=\mathcal{I}(t)\mathcal{Q}\tilde{\chi}(0)+\mathcal{K}(t)\mathcal{P}\tilde{\chi}(t), 
\label{solPchi}
\eeq
with the kernels again given by Eqs.~({\ref{IDefinition}}) and ({\ref{KDefinition}}).

\subsection{Removal of the Born Approximation and the Non-Markovian Regression Theorem}

Returning to Eq.~({\ref{derPExact}}) for the effective density operator, we 
now consider the inhomogeneous term $\mathcal{I}(\tau)\mathcal{Q}\tilde{\Upsilon}(t,0)$. 
If we were to make the Born approximation, and assume that the physical density operator factorises {at all times},  
$\chi(t)\approx\rho(t)\otimes\rho_R$, then $\mathcal{Q}\tilde{\Upsilon}(t,0)=0$ and the inhomogeneous term vanishes. 
Analogously, in Eq.~({\ref{solPchi}}) we see that in assuming factorising {initial conditions}, $\chi(0)\approx\rho(0)\otimes\rho_R$, 
the inhomogeneous term for the physical density operator vanishes. In these cases the equations of motion 
for the effective and the physical density operator become identical, i.e. we have 
$\partial_{t}\mathcal{P}\tilde{\chi}(t)=\mathcal{K}(t)\mathcal{P}\tilde{\chi}(t)$ and 
$\partial_{\tau}\mathcal{P}\TU=\mathcal{K}(\tau)\mathcal{P}\TU$. We conclude that we must make the Born approximation 
at all times for the standard regression theorem to apply. 

We now turn to the key insight of this work which allows us 
to remove the Born approximation. 
Since $B$ is a system operator, and assuming $[H_E,\rho_R]=0$, it can be shown 
that $\mathcal{Q}\tilde{\Upsilon}(t,0)=B \mathcal{U}(t)\mathcal{Q} \tilde{\chi}(t)$, where 
$\mathcal{U}_0(t)\tilde{\chi}(t)=\smash{U_0(t)\tilde{\chi}(t)U_0^{\dagger}(t)}$. 
The object $\mathcal{Q} \tilde{\chi}(t)$ represents deviations from factorability of 
the physical density operator. However, we already have an exact form for this, namely Eq.~({\ref{solQchi}}). 
Assuming factorising \emph{initial conditions only}, the second term in 
Eq.~({\ref{solQchi}}) is zero, and using what remains in Eq.~({\ref{derPExact}}) gives  
\beq
\partial_{\tau}\mathcal{P}\TU=\mathcal{I}'(t,\tau)\mathcal{P}\tilde{\chi}(t)+\mathcal{K}(\tau)\mathcal{P}\TU,
\label{derPExact2}
\eeq
where the new inhomogeneous term is given by 
$\mathcal{I}'(t,\tau)=\mathcal{I}(\tau)\mathcal{Q}B\mathcal{U}_0(t)(\openone-\Sigma(t))^{-1}\Sigma(t)\mathcal{P}$. 
Eq.~({\ref{derPExact2}}) is an exact equation of motion for the reduced effective density operator, 
in which the inhomogeneous term depends on the reduced physical density operator, 
which obeys the exact equation of motion Eq.~({\ref{solPchi}}) with $\mathcal{Q}\tilde{\chi}(0)=0$.

Though Eqs.~({\ref{derPExact2}}) and ({\ref{solPchi}}) are exact, calculating explicit forms for the kernels is 
difficult. The utility of the projection operator approach used here is that it allows for a systematic expansion in 
the system--environment coupling strength $\alpha$. 
Expanding the kernels appearing in Eq.~({\ref{derPExact2}}) to second order in $\alpha$ 
and moving back into the Schr\"{o}dinger picture we find  
\beq
\!\!\!\partial_{\tau}\Lambda(t,\tau)\!=i[\Lambda(t,\tau),H_S]+\mathcal{D}(\Lambda(t,\tau))+\mathcal{C}(\varrho(t,\tau)),
\label{derLambda}
\eeq
where the effective density operator enters through
\begin{align}
\label{D2ndOrder}
\!\!\!\mathcal{D}(\Lambda(t,\tau))=-\int_0^{\tau}\!\!\mathrm{d}s\mathrm{Tr}_E\big[H_I,\big[\tilde{H}_I(-s),\Lambda(t,\tau)\rho_R\big]\big],
\end{align}
and the physical density operator enters through
\begin{align}
\mathcal{C}(\varrho(t,&\tau))=\nonumber\\
-&\int_{\tau}^{\tau+t}\!\!\!\!\mathrm{d}s\mathrm{Tr}_E\big[H_I,\tilde{B}(-\tau)\big[\tilde{H}_I(-s),\varrho(t,\tau)\rho_R\big]\big],
\label{C2ndOrder}
\end{align}
with $\varrho(t,\tau)=U_S(\tau)\rho(t)U_S^{\dagger}(\tau)$, $\tilde{B}(-\tau)=U_S(\tau)BU_S^{\dagger}(\tau)$, and we have 
absorbed factors of $\alpha$ into the interaction Hamiltonians, i.e. $\alpha H_I\to H_I$. 
Let us review what approximations have been made. 
We assumed factorising \emph{initial conditions}, $\chi(0)=\rho(0)\otimes\rho_R$, and expanded the kernels to second order in the 
system-environment coupling strength. From this point onwards no further approximations are necessary. Finally, we note that 
$\rho(t)$ entering Eq.~({\ref{C2ndOrder}}) can be found at no additional cost since to the same level of approximation 
we have $\partial_t\rho(t)=-i[H_S,\rho(t)]+\mathcal{D}(\rho(t))$. 

Before proceeding, we note that we can obtain a time-independent equation of motion for $\Lambda(t,\tau)$ by making 
a Markovian approximation and let $\tau\to\infty$ in Eq.~({\ref{derLambda}}). We then find 
$\mathcal{C}(\varrho(t,\tau))=0$ and the inhomogeneous term disappears. 
In this case the regression theorem is recovered since $\rho(t)$ and $\Lambda(t,\tau)$ obey the same 
equation of motion. Recalling that we also find $\mathcal{C}(\varrho(t,\tau))=0$ when making the Born approximation, 
$\chi(t)\approx \rho(t)\otimes\rho_R$, we conclude that in the present context 
the Markovian approximation cannot be made without also implicitly making the Born approximation. 
Is the converse also true? 
Is it possible to not make the Markovian approximation by leaving the 
integration limit in Eq.~({\ref{derLambda}}) at $\tau$, yet at the same time make the Born approximation and neglect 
the inhomogeneous term $\mathcal{C}(\varrho(t,\tau))$? 
This is what one would obtain naively applying the regression theorem to a non-Markovian master equation for the physical density 
operator. In the following we will see that this approach is ill-advised and can give rise to unphysical results.

\section{Application to a Driven Semiconductor Quantum Dot Coupled to Acoustic Phonons}

We now use our results and consider a driven semiconductor QD in a non-Markovian acoustic 
phonon environment~\cite{ramsay10,mccutcheon10_2,mccutcheon13}. The QD is described by 
ground and single exciton states $\ket{g}$ and $\ket{e}$, and the laser by a constant Rabi frequency $\Omega$ and 
detuning $\delta$. 
In a rotating frame, and within the dipole and 
rotating wave approximations the Hamiltonian is given by $H=H_S+H_I+H_E$, with 
$H_S=\delta\sigma^{\dagger}\sigma+(\Omega/2)(\sigma^{\dagger}+\sigma)$, 
$H_I=\sigma^{\dagger}\sigma\sum_kg_k(b_k^{\dagger}+b_k)$ and 
$\smash{H_E=\sum_k \w_k b_k^{\dagger}b_k}$, where $\sigma=\ketbra{g}{e}$, a phonon with wave vector $k$ and 
frequency $\w_k$ is
described by creation and annihilation 
operators $b_k^{\dagger}$ and $b_k$, and we take a 
thermal state for the phonon environment $\rho_R=\exp[-H_E/k_B T]/\mathrm{Tr}[\exp[-H_E/k_B T]]$, with 
$T$ the sample temperature. 
The exciton--phonon interaction is characterised by coupling constants $g_k$, 
which ultimately enter only through the spectral density $J(\w)=\sum_k g_k^2\delta(\w-\w_k)$. 
For coupling to longitudinal acoustic phonons we can take the form $J(\w)=\eta\,\w^3\exp[-(\w/\w_c)^2]$, with $\eta$ the 
QD--phonon coupling strength, and $\w_c$ the cut-off frequency,  
whose inverse gives the memory time of 
the environment~\cite{mccutcheon10_2}. We tune the laser to the phonon-shifted QD transition frequency, 
$\delta=\int_0^{\infty}\mathrm{d}\w J(\w)/\w$, set $\Omega=0.12~\mathrm{ps}^{-1}$, 
and use the realistic parameters $\eta=0.03~\mathrm{ps}^2$ and $\w_c=2.2~\mathrm{ps}^{-1}$, 
with $T=4~\mathrm{K}$. 
The steady state first order correlation function of the QD emission is 
$g^{(1)}(\tau)=\lim_{t\to \infty}\av{\sigma^{\dagger}(t+\tau)\sigma(t)}$, which we calculate with 
Eq.~({\ref{derLambda}}), adding a term 
$\Gamma (\sigma\Lambda(t,\tau)\sigma^{\dagger}-\frac{1}{2}\{\sigma^{\dagger}\sigma,\Lambda(t,\tau)\})$ with $1/\Gamma=100~\mathrm{ps}$ to capture 
spontaneous emission.  Including spontaneous emission in this way assumes that the photonic environment is strictly Markovian, and 
is justified fully in the Appendix. Having obtained the first order correlation function the incoherent emission spectrum is defined as 
$S(\Delta\w)=\mathrm{Re}[\int_0^{\infty}\mathrm{d}\tau (g^{(1)}(\tau)-g^{(1)}(\infty))\e^{-i\Delta\w\tau}]$.

\begin{figure}
\begin{center}
\includegraphics[width=0.48\textwidth]{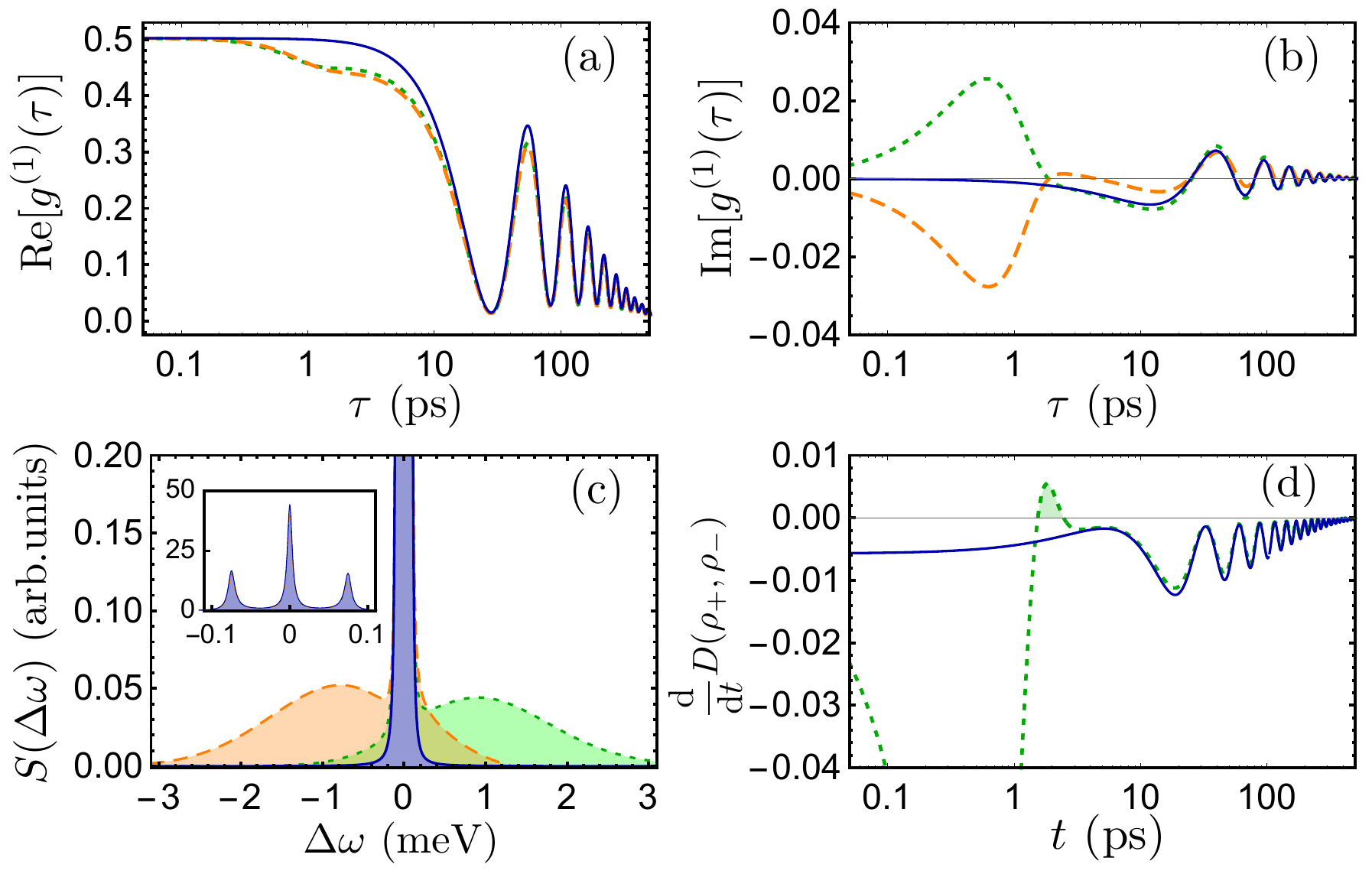}
\caption{Real (a) and imaginary (b) parts of the first order correlation function, calculated 
using a Markovian approximation (solid blue), the full non-Markovian theory (dashed orange), and the non-Markovian 
theory but neglecting the inhomogeneous term in Eq.~({\ref{C2ndOrder}}) (green dotted). Plot (c) shows the corresponding 
emission spectrum with the inset showing a different scale on which the Mollow triplet can be seen. 
From the main part of (c) it is seen that only the full non-Markovian theory 
correctly captures the phonon sideband at lower energies. Plot (d) shows the derivative of the trace 
distance between two states evolved from different initial conditions, whose positive 
values for times $\sim 1~\mathrm{ps}$ demonstrates backflow of information and true non-Markovianity.}
\label{AllPlots}
\end{center}
\end{figure}

Fig.~{\ref{AllPlots}} shows the real (a) and imaginary part (b) of $g^{(1)}(\tau)$ calculated using the Markovian approximation 
(taking $\tau\to\infty$ in Eqs.~({\ref{D2ndOrder}}) and ({\ref{C2ndOrder}}), solid blue), 
the full non-Markovian theory (dashed orange), and using the naive non-Markovian 
theory (neglecting the inhomogeneous term in Eq.~({\ref{C2ndOrder}}), dotted green). We 
see that for times less than the environment correlation time of $\sim 1~\mathrm{ps}$, 
all three theories predict quite distinct behaviour, reflecting 
the fact that non-Markovian effects are most important and these timescales. 
Plot (c) shows the corresponding incoherent emission 
spectrum, which on the inset scale displays the well-known Mollow triplet. 
From the main part of (c), we
see that the Markovian theory, which predicts no short time oscillations, correspondingly predicts no spectral features 
at large frequencies. The full non-Markovian theory, however, predicts a broad sideband at lower 
emission frequencies. 
This sideband is well-known 
experimentally~\cite{Matthiesen2013,Besombes2001,Favero2003,Ahn2005}, 
and is attributed to phonon emission, which our theory supports. 
Thus, the phonon sideband in the emission spectrum is a 
signature of non-Markovian behaviour. 
This is a key feature of this work; observation of non-Markovian behaviour of one-time expectation values 
typically necessitates initialising a system in a well-defined state and tracking dynamics 
on very short timescales ($\mathrm{ps}$ in this example). 
Steady-state two-time correlation functions, on the other hand, capture fluctuations of a system from equilibrium. 
Non-Markovian behaviour of these fluctuations can be much more readily observed 
since their Fourier transform corresponds to an emission 
spectrum~\footnote{We note that numerical investigations reveal that for certain parameters the 
non-Markovian theory can predict spectra which take on slightly negative values. 
It is believed that this signifies a limitation of the second-order approximation used, though may also point towards 
the need for a refined definition of the commonly used steady-state emission spectrum.}.

We note that while the phonon sideband has been  
calculated previously, it has so only in the zero driving limit $\Omega\to 0$ where 
the model becomes exactly solvable and the Mollow triplet is 
not present~\cite{Besombes2001,Favero2003,Ahn2005,Roy2015}. The theory 
presented here works for non-zero $\Omega$, allowing us to calculate the fraction 
of power emitted into the phonon sideband, which for the realistic parameters used here gives $\sim 10\%$, in good 
agreement with recent experiments~\cite{Matthiesen2013}.

Interestingly, it can be seen that the naive non-Markovian theory predicts a sideband at higher 
energies, in contrast to both intuition and experimental evidence. 
The inhomogeneous term in Eq.~({\ref{C2ndOrder}}) which the naive approach ignores 
captures deviations of the true state of the environment from the 
reference state $\rho_R$ used in the master equation. For the emission spectrum, these deviations are important, since 
we assumed $\rho_R$ to be a thermal state with respect to the QD ground state, which is not the correct 
initial condition for an emission process. 
This reveals why in neglecting the inhomogeneous term the sideband incorrectly appears 
at higher energies; since it assumes the environment to be in equilibrium with respect to the QD ground state, it 
inadvertently gives dynamics which correspond more to an absorption spectrum. 
We note that this correspondence is only approximate, and is not expected to be a general feature.

The steady-state correlation function we have calculated captures fluctuations of the QD about its steady state, 
and our results suggest these fluctuations are non-Markovian in nature. 
In order for this is be confirmed, we calculate a non-Markovianity witness in the form 
of the derivative of the trace distance 
$D(\rho_+,\rho_-)=\frac{1}{2}|\rho_+(t)-\rho_-(t)|$, where 
$\rho_+(t)$ and $\rho_-(t)$ are physical density operator states evolved from 
two different initial states $\rho_{\pm}(0)=\frac{1}{2}(\openone\pm\sigma_y)$ 
with $\sigma_y=-i\ketbra{e}{g}+i\ketbra{g}{e}$~\cite{breuer09}. 
We are interested here in the evolution 
of reduced physical density operators since these characterise the behaviour of physical QD exciton, 
and as such use the equation of motion $\partial_t \rho(t)=-i[H_S,\rho(t)]+\mathcal{D}(\rho(t))$ (i.e. without 
inhomogeneous term). 
A positive derivative of the trace distance is interpreted as a flow of information from the environment 
into the system, and is a sufficient condition to prove indivisibility of the underlying dynamical map, both 
of which can be considered definitions of non-Markovianity~\cite{breuer09,rivas10,Guarnieri2014}. 
In Fig.~{\ref{AllPlots}}(d) we show $\frac{\mathrm{d}}{\mathrm{d}t}D(\rho_+,\rho_-)$ calculated 
using the non-Markovian theory (dotted, green), and within the Markovian approximation (solid, blue). 
We see that our non-Markovian theory gives rise to a time interval during which the derivative 
is positive, confirming true non-Markovian behaviour. 

\section{Summary}

We have developed an extension to the quantum regression theorem, valid to second order in the system-environment 
coupling strength, and invoking the Born approximation at a single fixed initial time. These results have been used 
to demonstrate that phonon sidebands in the resonance fluorescence 
emission spectra of a QD are a signature of non-Markovian behaviour. In this context, 
it was shown that this non-Markovian behaviour is associated with a flow of information 
from the phonon environment back into the QD exictonic system, which is a sufficient condition 
to prove indivisibility of the underlying dynamical map. 
The projection operator method used here is an ideal starting point to include higher order system--environment 
coupling terms, which can in some cases lead to an exact resummation~\cite{barnes12}. 
Finally, it will be interesting to investigate how 
the results obtained here 
can be used to optically quantity 
non-Markovian behaviour~\cite{Wolf2008,breuer09,rivas10,Guarnieri2014,Luoma2014}.

\appendix
\section{Effective density operator master equation for time-dependent interaction Hamiltonians} 

Here we give an extension to the results provided in the main text which facilitates the inclusion 
of time-dependent interaction Hamiltonians. For a time-dependent interaction Hamiltonian 
we can write the complete Schr\"{o}diner picture Hamiltonian 
in the form $H(t)=H_S+\alpha H_I(t)+H_E$. In this case defining an interaction picture proceeds analogously 
as in the main text, and the interaction picture equation of motion for the effective density operator again 
takes the form of Eq.~({\ref{LDefinition}}), though now we have 
\beq
\tilde{H}_I(\tau)=U_0^{\dagger}(\tau) H_I(\tau) U_0(\tau),
\label{HIint}
\eeq
with $H_I(\tau)$ the Schr\"{o}dinger picture interaction Hamiltonian at time $\tau$, 
and $\TU=U_0^{\dagger}(\tau)U(t+\tau,t) [B\chi(t)] U^{\dagger}(t+\tau,t)U_0(\tau)$ where  
the time evolution operator satisfies $i\partial_t U(t,t_0) = H(t) U(t,t_0)$ with $U(t_0,t_0)=\openone$. 
For this time-dependent interaction Hamiltonian the derivation of the effective density operator 
master equation proceeds precisely as in the main text, 
and we again arrive at the general expression in Eq.~({\ref{derPExact2}}), the only difference being that 
the implicit occurrences of the interaction Hamiltonians are defined through Eq.~({\ref{HIint}}). 
Expanding to second order in the system-environment coupling strength proceeds analogously, 
though some care must be taken when moving back into the Schr\"{o}dinger picture. 
For a time-dependent Hamiltonian the Schr\"{o}dinger picture equation of 
motion for the effective density again has the form   
$\partial_{\tau}\Lambda(t,\tau)=-i[H_S,\Lambda(t,\tau)]+\mathcal{D}(\Lambda(t,\tau))+\mathcal{C}(\varrho(t,\tau))$, though now  
\begin{align}
&\mathcal{D}(\Lambda(t,\tau))=\nonumber\\
-&\int_0^{\tau}\!\!\mathrm{d}s\mathrm{Tr}_E\big[\tilde{H}_I(\tau,0),\big[\tilde{H}_I(\tau-s,-s),\Lambda(t,\tau)\rho_R\big]\big],\label{Dt}
\end{align}
and the inhomogeneous term is given by 
\begin{align}
\!\!\mathcal{C}(\rho(t))=-\!\int_{\tau}^{\tau+t}\!\!\!\!\mathrm{d}s&\mathrm{Tr}_E\Big[\tilde{H}_I(\tau,0),\tilde{B}(-\tau)\nonumber\\
&\big[\tilde{H}_I(t+\tau-s,-s),\varrho(t,\tau)\rho_R\big]\Big]\label{Ct},
\end{align}
and we have defined $\tilde{H}_I(t_1,t_2)=U_0^{\dagger}(t_2) H_I(t_1) U_0(t_2)$. Note that in order to recover the case 
for a time-independent interaction Hamiltonian we simply set the first time argument in $\tilde{H}_I(t_1,t_2)$ to zero.

\section{Inclusion of Spontaneous Emission within the Markovian Approximation}

Here we give details of how spontaneous emission can be included into the effective density operator 
master equation in the context of the quantum dot (QD) example in the main text. 
To so so we consider an optically driven QD coupled to both a phonon and photon reservoir. Within the dipole 
and rotating wave approximations the total Schr\"{o}dinger picture Hamiltonian in a frame rotating 
at the laser frequency $\w_l$ takes the form $H(t)=H_S+H_{I1}+H_{I2}(t)+H_{E1}+H_{E2}$ 
where $H_S=\delta\sigma^{\dagger}\sigma+(\Omega/2)(\sigma^{\dagger}+\sigma)$, 
$H_{I1}=\sigma^{\dagger}\sigma\sum_k g_k(b_k^{\dagger}+b_k)$, $H_{E1}=\sum_k \w_k b_k^{\dagger} b_k$, 
while 
\begin{align}
H_{I2}(t)=\sum_q h_q (\sigma a_q^{\dagger}\e^{-i \w_l t}+\sigma^{\dagger} a_q \e^{i \w_l t}),
\end{align}
and $H_{E2}=\sum_q \nu_q a_q^{\dagger} a_q$, 
where parameters with a $k$ subscript refer to phonons, while $h_q$ is the coupling constant between the quantum dot 
and photonic mode $q$, described by creation operator $a_q^{\dagger}$ and frequency $\nu_q$. 
Since the total interaction Hamiltonian $H_I(t)=H_{I1}+H_{I2}(t)$ is time-dependent we must use 
Eqs.~({\ref{Dt}}) and ({\ref{Ct}}), where the trace is now taken over both phonon and photon degrees of freedom. 
Assuming that $H_{I1}$ and $H_{I2}(t)$ contain no environment operators that act in the same Hilbert space (as is the case 
in our example), one finds that provided $\mathrm{Tr}_{E1}[H_{I1}\rho_R]=0$, the `cross' terms mixing 
$H_{I1}$ and $H_{I2}(t)$ in Eqs.~({\ref{Dt}}) and ({\ref{Ct}}) vanish, and we can write 
$\partial_{\tau}\Lambda(t,\tau)=-i[H_S,\Lambda(t,\tau)]
+\mathcal{D}_1(\Lambda(t,\tau))+\mathcal{C}_1(\varrho(t,\tau))
+\mathcal{D}_2(\Lambda(t,\tau))+\mathcal{C}_2(\varrho(t,\tau))$, 
where $\mathcal{D}_1$ and $\mathcal{C}_1$ contain only phonon terms, i.e. they 
are Eqs.~({\ref{Dt}}) and ({\ref{Ct}}) with $\tilde{H}_I(t_1,t_2)\to U_0^{\dagger}(t_2) H_{I1} U_0(t_2)$, 
and $\mathcal{D}_2$ and $\mathcal{C}_2$ contain only photon terms, 
i.e. Eqs.~({\ref{Dt}}) and ({\ref{Ct}}) with $\tilde{H}_I(t_1,t_2)\to U_0^{\dagger}(t_2) H_{I2}(t_1) U_0(t_2)$. 
As in the main text we have $U_0(t)=U_S(t)U_E(t)$ though now $U_E(t)=\exp[- i (H_{E1}+H_{E2})t]$.

Let us consider the term in $\mathcal{D}_2(\Lambda(t,\tau))$ in more detail. The relevant interaction Hamiltonian 
can be written 
\beq
\tilde{H}_{I2}(\tau-s,-s)=\tilde{\sigma}(-s) \tilde{{A}}^{\dagger}(-s)\e^{-i \w_l (\tau-s)}+\mathrm{h.c.},
\eeq
where $\tilde{\sigma}(-s)=\e^{-i H_S s} \sigma \e^{i H_S s}$ and 
$\tilde{A}(-s)=\sum_q h_q a_q \e^{i \nu_q s}$. Assuming a zero temperature thermal 
state environment for the photons, i.e. 
$\rho_R = \rho_{R1}\rho_{R2}$ with $\rho_{R1}$ the state of the phonon environment and 
$\rho_{R2}=\exp[-\beta \sum_q \nu_q a_q^{\dagger} a_q]/\mathrm{Tr}[\exp[-\beta \sum_q \nu_q a_q^{\dagger} a_q]]$ 
with $\beta\to\infty$, we find 
$\mathrm{Tr}_E[A^{\dagger} \tilde{A}^{\dagger}(-s) \rho_R]=\mathrm{Tr}_E[A \tilde{A}(-s) \rho_R]=\mathrm{Tr}_E[A^{\dagger} \tilde{A}(-s) \rho_R]=0$, 
and we are left with 
\begin{align}
\mathcal{D}_2(\Lambda(t,\tau))&=-\int_0^{\tau}\mathrm{d}s\mathrm{Tr}_E[A \tilde{A}^{\dagger}(-s) \rho_R]\e^{i\w_l s}\nonumber\\
&\Big(\sigma^{\dagger} \tilde{\sigma}(-s) \Lambda(t,\tau)-\tilde{\sigma}(-s)\Lambda(t,\tau)\sigma^{\dagger}\Big)+\mathrm{h.c.}
\end{align}
We now make a Markovian approximation, with respect to the photon environment only, and approximate 
the remaining correlation function as a delta-function, i.e. we take 
$\mathrm{Tr}_E[A \tilde{A}^{\dagger}(-s) \rho_R]=\zeta \delta(s)$, in which case we find
\beq
\mathcal{D}_2(\Lambda(t,\tau))=\Gamma\Big(\sigma\Lambda(t,\tau)\sigma^{\dagger}-\frac{1}{2}\{\sigma^{\dagger}\sigma,\Lambda(t,\tau)\}\Big),
\label{GammaTerm}
\eeq
where $\Gamma = 2 \alpha^2 \zeta$ is the spontaneous emission rate. Considering now the 
photonic inhomogeneous term, $\mathcal{C}_2(\varrho(t,\tau))$, making the same Markovian 
approximation for a zero temperature environment results in $\mathcal{C}_2(\varrho(t,\tau)=0$ for 
all times $\tau>0$ of interest owing to the integration limits in Eq.~({\ref{Ct}}).
As such, within the Markovian approximation for the photonic environment, we can 
simply neglect the photon terms at a Hamiltonian level, provided we add a 
term equal to Eq.~({\ref{GammaTerm}}) to the equation of motion Eq.~({\ref{derLambda}}) in the main text. 
We note that approximating the photonic correlation functions as delta-functions is expected 
to be a good approximation for quantum dots in free space or in low Q-factor cavities, where 
photon correlation times of $\sim 10^{-2}-10^{-3}~\mathrm{ps}$ are typically orders of magnitude 
shorter than the phonon bath correlation time of $\sim1~\mathrm{ps}$~\cite{mccutcheon13,Roy-choudhury2015}.

\begin{acknowledgments}
I would like to thank Jesper M{\o}rk, Ahsan Nazir and Jake Iles-Smith for useful discussions. 
This work was funded by project SIQUTE (contract EXL02) of the European Metrology Research Programme (EMRP). 
The EMRP is jointly funded by the EMRP participating countries within EURAMET and the European Union.
\end{acknowledgments}


%

\end{document}